\documentclass[12pt]{article}
\usepackage{amsmath,amssymb}
\begin{document}

\begin{titlepage}

\setlength{\baselineskip}{18pt}

                            \begin{center}

                            \vspace*{2cm}

        \large\bf  Rate of parity violation from measure concentration\\

                            \vfill

              \small\sf NIKOS \  KALOGEROPOULOS\\

                            \vspace{0.2cm}

 \small\sf  Department of Science\\
            BMCC-The City University of New York\\
            199 Chambers St., New York, NY 10007, USA\\
                            \end{center}

                            \vfill

                     \centerline{\normalsize\bf Abstract}
                            \vspace{1mm}

\normalsize\rm\noindent\setlength{\baselineskip}{18pt} We present
a geometric argument determining the kinematic (phase-space)
factor contributing to the relative rate at which degrees of
freedom of one chirality come to dominate over degrees of freedom
of opposite chirality, in models with parity violation. We rely on
the measure concentration of a subset of a Euclidean cube which is
controlled by an isoperimetric inequality. We provide an
interpretation of this result in terms
of ideas of Statistical Mechanics. \\

                             \vfill


                             \vfill

\noindent\rule{7.5cm}{0.2mm}\\
\begin{tabular}{ll}
\noindent\small\rm E-mail: & \small\rm nkalogeropoulos@bmcc.cuny.edu\\
                  & \small\rm nkaloger@yahoo.com
\end{tabular}

\end{titlepage}


                                 \newpage

\setlength{\baselineskip}{18pt}

                    \centerline{\sc 1. \  Introduction}

                               \vspace{7mm}

The rate at which a system with many degrees of freedom approaches
its equilibrium state [1],[2] is one of its most important
characteristics, ranking alongside the existence and the
determination of the equilibrium configuration itself. Arguments
determining such rates have appeared since the earliest days of
statistical mechanics [1] and have also played an important role
in classical and quantum field theories [3],[4]. There is a far
greater variety of non-equilibrium than equilibrium phenomena, in
which even conventional thermodynamic quantities and relations may
need redefinition [5]-[8]. For such a reason alone, it may be of
some importance to be able to determine rates of convergence
to equilibrium in an, as much as possible, model-independent way.
In this paper we attempt to determine such a rate of convergence
for a system exhibiting parity violation.\\

Parity violation is one of the most intriguing, experimentally
verified effects encountered in physics. This phenomenon has been
a subject of extensive studies due to its great importance for
particle physics [9]-[13], nuclear physics [14]-[16], quantum
gravity [17]-[19], astrophysics [20] and atomic physics [21].\\

In many instances the detailed mechanism of parity violation is
not understood [9]-[13] at a fundamental level. There are other
cases, effective field theories being an example, where it may not
even be important to know the fundamental underpinnings of such a
behavior. In either case, conclusions relying on general kinematic
grounds, which are largely independent of the specific
interactions of the model at hand, may be of interest in
determining general characteristics of the behavior of the degrees
of freedom under consideration.\\

An outline of this paper is as follows: In Section 2, we construct
a simple geometric model that quantifies phase-space contribution
to the rate with which the asymmetric production of the two
chirality degrees of freedom leads to a dominance of the one kind
over the other. We find that this factor of the rate of dominance
has a Gaussian behavior, when the number of degrees of freedom \
$n$ \ is very large. In Section 3, we provide a physical
explanation for this emergent Gaussian behavior by associating it
to the rate at which a classical statistical system approaches its
equilibrium state. In Section 4, we draw some conclusions, provide
the context and make some additional
comments regarding the applicability of this approach.\\

                              \vfill


 \centerline{\sc 2. \ A geometric model and Gaussian concentration}

                             \vspace{3mm}

We assume from the outset the existence of a mechanism of parity
violation without attempting to provide any details of its origin
or its specific workings. \\

Consider a field \ $\psi$ \ with a definite chirality. Then \
$\psi$ \ is an eigenfunction of the chirality operator with
eigenvalues \ $\pm 1$. \ In particle physics [9], for instance,
the chirality of space-time fermionic degrees of freedom is
determined by the eigenvalue of \ $\psi$ \ under the action of \
$\frac{1}{2}(1 \pm\gamma_5)$, \ as
\begin{equation}
           \frac{1 \pm \gamma_5}{2} \  \psi \ = \ \pm \ \psi
\end{equation}
where \ $\gamma_5$ \ denotes the top element of the Clifford
algebra of gamma matrices [9]. We indicate the chirality
eigen-states of \ $\psi$ \ by \ $|+\rangle$ \ and \ $|-\rangle$ \
in analogy with the spin-$\frac{1}{2}$ \ notation. Let the field
theoretical model have \ $n$ \ such degrees of freedom. Eventually
\ $n$ \ should be considered very large, namely, we will be
interested in the ``thermodynamic" limit \ $n\rightarrow\infty$. \
The possible chirality eigen-states form a basis of a tensor
product representation of \ $\mathbb{Z}_2^{\otimes n}$, \ where \
$\mathbb{Z}_2 = \{-1, +1 \}$ \ and as such they have the form
\begin{equation}
     |p_1\rangle \otimes |p_2\rangle \otimes \cdots \otimes
     |p_n\rangle
\end{equation}
where \ $p_i, \ i=1,2,\ldots,n$ \ stand for \ $+$ \ or \ $-$ in
this equation and for the rest of this paper. The carrier space of
such a representation is the Hilbert space of the $n$-particle
states \ $\mathcal{H}_n$. \ Let us focus exclusively on the
chirality behavior of the degrees of freedom \ $\psi$ \ and forget
any other properties that they may possess. Then \ $\mathcal{H}_n$
\ has a basis comprised of elements of the form (2).
Geometrically, \ $\mathcal{H}_n$ \ can be seen as the set of
vertices of the Euclidean cube \ $\mathbb{E}_2^n$ \ in \
$\mathbb{R}^n$ \ whose
side has a length of \ $2$ \ units. This cube is the phase space
pertinent to the parity degrees of freedom of the system.\\

Consider now two $n$-particle chirality eigen-states \
  $|P \rangle \ = \ |p_1\rangle \otimes |p_2\rangle \otimes
  \cdots \otimes |p_n\rangle$ \ and \
  $|P'\rangle \ = \ |p_1'\rangle \otimes |p_2'\rangle \otimes
    \cdots \otimes |p_n'\rangle$. \ These two eigen-states
can be geometrically represented as two vertices of the cube \
$\mathbb{E}_2^n$. \ We would like to find a reasonable expression
for a ``distance" between such states/vertices. This is equivalent
to choosing a metric for \ $\mathbb{E}_2^n$. \ One reasonable
choice would be the induced metric on \ $\mathbb{E}_2^n$ \ from
the Euclidean metric of \ $\mathbb{R}^n$. \ A potential drawback
of such a choice, which is induced by the \ $l^2$ \ norm, is that
it does not only express intrinsic properties of \
$\mathbb{E}_2^n$ \ but also reflects, to some extent, the metric
embedding \ $\mathbb{E}_2^n \hookrightarrow \mathbb{R}^n$. \ For
this reason, as well as for greater simplicity, we prefer to avoid
such a choice, and use instead the normalized Hamming
 metric \
 $d_n: \mathbb{E}_2^n \times \mathbb{E}_2^n \rightarrow \mathbb{R}_+$ \ instead,
 which is defined by
 \begin{equation}
         d_n(|P\rangle, |P'\rangle) \ = \ \frac{1}{2n} \ \sum_{i=1}^n
         |p_i-p_i'|
 \end{equation}
where \ $|\cdot|$ \ indicates the absolute value in \
$\mathbb{R}$. \ This choice corresponds to the choice of an \
$l^1$ \ norm for \ $\mathbb{E}_2^n$ \ up to a normalizing constant
\ $1/2n$. \ It is easy to check that \ $d_n$ \ satisfies all three
properties (positivity, symmetry and the triangular inequality)
required of metrics. We can define, following the spirit of (3),
the distance between a vertex \ $|x\rangle \in\mathbb{E}_2^n$ \
and the set \ $A\subseteq\mathbb{E}_2^n$ \ as
\begin{equation}
        d_n(|x\rangle, A) \ = \ \inf \{ d_n(|x\rangle, |P\rangle),
            \ \ \forall \ |P\rangle \in \mathbb{E}_2^n \}
\end{equation}
Continuing in this mode, we define the \ $\epsilon$-extension of \
$A$ \ [22], [23] as
\begin{equation}
    A_{\epsilon} \ = \ \{ |x\rangle \in \mathbb{E}_2^n :
                       d_n(|x\rangle, A) \ \leq \epsilon, \ \ \epsilon > 0 \}
\end{equation}
Then \ $A_{\epsilon}$ \ is the set of all points (vertices) of \
$\mathbb{E}_2^n$ \ which are at a Hamming distance at most \
$\epsilon$ \ away from \ $A$. \ Since \ $\mathbb{E}_2^n$ \ is a
discrete set, it is reasonable to consider as the measure (the
``volume") of \ $A\subseteq \mathbb{E}_2^n$ \ its cardinality \
$\mathrm{card}(A)$. \ We will eventually be interested in a
comparison of the number of states (vertices) of the phase space \
$\mathbb{E}_2^n$, \ representing the dominant chirality, with the
total number of states (vertices) of \ $\mathbb{E}_2^n$. \ In
light of this future comparison, it is more advantageous to work
with the normalized cardinality \ $P(A)$ \ of \ $A$ \ instead,
defined as
\begin{equation}
    P(A) \ = \ \frac{\mathrm{card}(A)}{2^n}
\end{equation}
For concreteness, we can assume, without any loss of generality,
that \ $A$ \ is a set of vertices of \ $\mathbb{E}_2^n$ \ which
represent states with more (or equal to) positive than negative
chirality degrees of freedom. Obviously \ $P(A)\geq \frac{1}{2}.$
\ For such \ $A$, \ [22], [23] defines the concentration function
\begin{equation}
h(A, \epsilon) \ = \ 1 - \inf \{P(A_{\epsilon}), \
A\subseteq\mathbb{E}_2^n,
                           \ \epsilon > 0 \}
\end{equation}
which measures the maximal fraction of the ``volume" of \
$\mathbb{E}_2^n$ \ not belonging to \ $A_{\epsilon}$. \ In our
particular case, \ $\epsilon$ \ is proportional to the difference
of the number of states in phase space of the dominant minus the
recessive chirality. It was proved in [24], [25] by relying on an
isoperimetric inequality on graphs, that
\begin{equation}
    h(A, \epsilon) \ \leq \ \frac{1}{2} \ e^{-2\epsilon^2n}
\end{equation}
This result can be interpreted as stating that, as the number of
degrees of freedom \ $n$ \ increases without any upper bound, more
states of the phase space \ $\mathbb{E}_2^n$ \ are concentrating
closer and closer to \ $A_{\epsilon}$. \ This rate of
concentration is given approximately  by \ $\epsilon \sim
n^{-\frac{1}{2}}$. \ This argument also implies that \
$A_{\epsilon}$ \ will eventually contain almost all the vertices
of \ $\mathbb{E}_2^n$ \ as \ $n\rightarrow\infty.$ \  Therefore,
chirality eigen-states containing more positive than negative
chirality degrees of freedom dominate so fast, that the recessive
chirality eigen-states disappear exponentially fast with \ $n$. \
The concentration function (7) contributes an exponential factor
to the rate of dominance of one chirality over the other. Another
factor to such a rate will be contributed by the specific dynamics
of the model at hand. Such dynamics will be manifested through the
paths, expressing the time evolution, connecting the vertices that
the system will occupy in the phase space \ $\mathbb{E}_2^n$. \ We
also observe in (8) that the concentration function is Gaussian
with respect to \ $\epsilon$. \ Gaussian functions are encountered
very frequently as a result of the law of large numbers: the
present result is no exception. What we have just found is a
geometric incarnation, in our specific context of the law of large
numbers [26]. \\

It is probably worth mentioning at this point, that, using the
theory of large deviations, one can arrive, through a different
line of reasoning, to the following upper bound for the
concentration function [22], [23]
\begin{equation}
   h(A, \epsilon) \ \leq \ 2 \ e^{-\frac{\epsilon^2n}{16}}
\end{equation}
In neither of these cases is the upper bound claimed to be
optimal. Roughly speaking, there is very little difference between
(8) and (9). What really matters the most is the convergence rate
of \ $h(A, \epsilon)$ \ in terms of \ $\epsilon^2n$, \ regardless
of its pre-factor in (8) or (9).\\

 The convergence behavior of \
$h(A, \epsilon)$, \ as \ $n\rightarrow\infty$, \ is physically
important because as does not depend on the value of the
production rate or on the specifics of the interactions of the
degrees of freedom of a given chirality. This is to be expected of
course, as this factor is due only to kinematics (phase space) and
ignores the specific dynamics of the model. As long as the degrees
of freedom of one chirality are produced at a higher rate than the
other, with the production rate difference \ $\epsilon$ \ being
constant in time, one chirality will asymptotically dominate over
the other at a rate which has a Gaussian factor involving the
excess \ $\epsilon$. \ We also assume that \ $\epsilon$ \ should
not approach zero faster than \
$n^{-\frac{1}{2}}$, \ if  results (8) or (9) are to hold.\\

                        \vspace{5mm}


  \centerline{\sc 3. \ On the origin of the Gaussian behavior}

                        \vspace{3mm}

The emergence of a Gaussian dependence of \ $h(A, \epsilon)$ \ in
terms of \ $\epsilon$ \ may be unexpected, so in this Section we
attempt to elucidate it. This procedure has two stages. Both of
them are inspired and follow the spirit of the analysis of
Boltzmann on the derivation of the thermodynamic properties of an
ideal gas [27]. We also stress the underlying geometric structure
as much as possible.\\

First, consider the unit sphere \ $S^n$. \ Its volume is given
[28] by
\begin{equation}
    vol (S^n) \ = \
    \frac{2\pi^{\frac{n+1}{2}}}{\Gamma(\frac{n+1}{2})}
\end{equation}
Using the Stirling approximation [29], for \ $n\rightarrow\infty$
\begin{equation}
\Gamma(n) \ =  \ \sqrt{\frac{2\pi}{n}} \
 \left(\frac{n}{e}\right)^n
\end{equation}
we can see that the asymptotic behavior of \ $vol(S^n)$ \ is given
by
\begin{equation}
       vol (S^n) \ \sim  \ (n+1)^{-\frac{n+1}{2}}
\end{equation}
The cube \ $\mathbb{E}_2^{n+1}$ \ with the normalized Hamming
distance (3) has diameter 1. To draw a clearer analogy with the
case of $S^n$, we consider, instead, the comparison with a
Euclidean cube \ $\mathbb{E}^n$ \ of side $2$ units. To be able to
make a reasonable comparison with \ $S^n$, \ we re-scale the
diameter (length of longest diagonal) of $\mathbb{E}^{n+1}$ to $1$
unit, which amounts to a re-scaling of its side length to
$(n+1)^{-\frac{1}{2}}$. Then
\begin{equation}
       vol (\mathbb{E}^{n+1}) \ = \ (n+1)^{-\frac{n+1}{2}}
\end{equation}
We observe that the leading term in the asymptotic behavior of the
volumes of \ $S^n$ \ and \ $\mathbb{E}^{n+1}$ \ is the same, as \
$n\rightarrow\infty$. \ Therefore in this approximation, we can
use \ $S^n$ \ instead of \ $\mathbb{E}^{n+1}$ \ in arguments
involving volumes, as is done in the next paragraphs.\\

The substitution of  \ $\mathbb{E}_2^n$ \ endowed with the Hamming
metric by \ $\mathbb{E}^n$ \ endowed with the Euclidean metric,
which was performed in the preceding paragraph can also be
justified more formally, without altering the essence of the
argument above, as follows: We can look at the ``dual description"
of the two cubes we compare, in terms of the behavior of the set
of real functions that are defined on them. Let \
$h:\mathbb{R}^n\rightarrow \mathbb{R}$. We note that these cubes
are also the unit spheres with respect to the usual \ $l^1$ \ and
\ $l^2$ \ norms. The corresponding norms on the spaces of
Lebesgue-integrable functions \ $L^1$ \ and \ $L^2$ \ of these
cubes are denoted by \ $\|h \|_1$ \ and \ $\|h\|_2$, \
respectively. Almost all functions encountered in particle physics
are analytic, at least in part of their domain of definition, so
the Taylor series expansion of such functions receives its
dominant contribution from their linear term, as long as the
function is not expanded around one of its critical points. For
such linear approximations of \ $h$, \ a combination of the
H\"{o}lder and the Kahane-Khinchine inequalities gives [30]
\begin{displaymath}
   \|h\|_1 \ \leq \ \|h\|_2 \ \leq \ \sqrt{2} \ \|h\|_1
\end{displaymath}
Therefore, the two cubes that we are considering with the
Euclidean and the Hamming metrics are quasi-isometric. So, if we
look at their metric properties, they are essentially the same,
``very roughly speaking". Since our explanation of the emergence
of the Gaussian behavior is only very rough, in the sense that as
\ $n\rightarrow\infty$, \ deviations of order \ $n^\frac{1}{2}$ \
or smaller  are inconsequential, as seen from
(8),(9), such a quasi-isometric equivalence is sufficient for our purposes.\\

As a second step we focus, as in the previous paragraph, on the
dual description [22], [23]: we consider the set of all
continuous, square-integrable functions \ $f_i: S^n \rightarrow
\mathbb{R}, \ i=1,\ldots, n+1$. \ Most of the physical functions
of interest are assumed to be smooth enough, and they satisfy
these conditions. Actually, this type of argument has been
developed for Lipschitz functions [26]. We should consider, once
more, \ $S^n$ \ as embedded in $\mathbb{R}^{n+1}$. \ Let \ $x_i
\in \mathbb{R}, \ i=1,2,\ldots, n+1$ \ be the Cartesian
coordinates in $\mathbb{R}^{n+1}$ \ and \
$r^2=\sum_{i=1}^{n+1}x_i^2$ \ indicate the radial distance from
the origin of this coordinate system. Let \ $dvol$ \ stand for the
Riemannian volume element of \ $S^n$ \ with respect to the round
metric. By using the well-known facts [29]
\begin{equation}
   \frac{1}{\sqrt{2\pi}} \int_{-\infty}^{+\infty}
   e^{-\frac{t^2}{2}} \ dt = 1
\end{equation}
and
\begin{equation}
   \prod_{i=1}^{n+1}  dx_i \ = \ r^n \ dr \ dvol
\end{equation}
we find
\begin{equation}
   \int_0^{\infty} e^{-\frac{r^2}{2}}r^n \ dr = \frac{
   (2\pi)^\frac{n+1}{2}}{vol(S^n)}
\end{equation}
We then get
\begin{equation}
  \int_{S^n} \| \sum_{i=1}^{n+1} x_if_i \|^2 \ dvol \ = \
     \frac{vol(S^n)}{(2\pi)^{\frac{n+1}{2}}}
     \int_{S^n} \| \sum_{i=1}^{n+1} x_if_i \|^2 \ dvol \
     \int_0^\infty r^{n}e^{-\frac{r^2}{2}} \ dr
\end{equation}
which can be rewritten as
\begin{equation}
   \frac{1}{vol(S^n)} \int_{S^n} \| \sum_{i=1}^{n+1} x_if_i \|^2 \ dvol \
  = \ \frac{1}{(2\pi)^{\frac{n+1}{2}}} \int_{\mathbb{R}^{n+1}}
    \| \sum_{i=1}^{n+1} x_if_i \|^2 \ e^{-\sum\limits_{i=1}^{n+1}
    \frac{x_i^2}{2}} \ dx_1 \cdot \ldots \cdot dx_{n+1}
\end{equation}
which eventually gives
\begin{equation}
 \frac{1}{n+1} \ \langle \ \| \sum_{i=1}^{n+1} s_i f_i \|^2 \ \rangle \ =
   \ \frac{1}{vol(S^n)} \int_{S^n} \| \sum_{i=1}^{n+1} x_if_i \|^2 \ dvol
\end{equation}
 where \ $s_i, \ i=1, \ldots, n+1$ \ are
Gaussian random variables with mean $0$ and variance \ $\sigma =
1$. \ Here \ $\langle \cdot \rangle$ \ denotes the Gaussian
average of its argument over \ $\mathbb{R}^{n+1}$, \ hence the
appearance of the \ $n+1$ \ in the denominator. The conclusion is
that averages of continuous, square integrable functions \ $f_i,
i=1,\ldots, n+1$ \ over \ $S^n$ \ can be expressed as Gaussian
averages  of these functions over \ $\mathbb{R}^{n+1}$ \ [22],[23],[26],[27].\\

The previous paragraph applies, to a great extent, for functions
on \ $S^n$. \ However, if one is not interested in the very
``small-scale" details of the geometry of \ $S^n$, \ then one may
choose to ``probe" \ $S^n$ \ with objects providing a
``resolution" of order at most \ $\epsilon$. \ The smallest
distances \ $x$ \ that we want to be able to distinguish in this
argument are of order of magnitude \ $n^{-\frac{1}{2}}$, \ which
is the approximate length of the side of the cube \
$\mathbb{E}^{n+1}$, \ having the same diameter as \ $S^n$. \
Unambiguous results can be drawn as long as \ $\epsilon = \sigma
\cdot \frac{1}{\sqrt{n}}$, \ which amounts to \ $\epsilon \sqrt{n}
= \sigma$. \ Then \ $S^n$ \ and \ $\mathbb{E}^{n+1}$ \ do not only
have the same volume (as $n\rightarrow\infty$), but also appear,
roughly, metrically the same at a scale of \ $x\sim \sigma$ \ or
larger. Within such a minimum length scale rough approximation,
the previous analysis of the Gaussian behavior of functions on \
$S^n$ \ applies equally well to \ $\mathbb{E}^{n+1}$. \ So
\begin{equation}
   \frac{1}{n+1} \ \langle \ \| \sum_{i=1}^{n+1} s_i f_i \|^2 \rangle \ =
   \ \int_{\mathbb{E}^{n+1}} \| \sum_{i=1}^{n+1} x_if_i \|^2 \
   d\mu_{\mathbb{E}^{n+1}}
\end{equation}
where \ $\| \cdot \|$ \ stands for the \ $L^2$ \ norm in the space
of interest in (17)-(20) and  \ $s_i$ \ has the form
\begin{equation}
         s_i \sim e^{-\epsilon^2n}
\end{equation}
which agrees, approximately, with the results of (8) and (9).\\

It is worth noticing, at this point, that there is nothing really
stochastic about the chirality eigen-states. The corresponding
chirality eigen-values are \ $\pm 1$ \ with probability \ $1$. \
This deterministic behavior can  be formally recovered in the
limit of shrinking the ``resolution" \ $\epsilon\rightarrow 0$, \
which amounts to re-scaling the variance of the Gaussian variables
\ $s_i$, \ so that it approaches zero, since
\begin{equation}
         \delta (x) \ = \ \lim_{\sigma\rightarrow 0}
         \frac{1}{\sqrt{2\pi}\sigma} \ e^{-\frac{x^2}{2\sigma^2}}
\end{equation}


                          \vspace{5mm}

          \centerline{\sc 4. \ Discussion and Conclusions}

                         \vspace{3mm}

In this work we presented a geometric argument determining the
phase-space contribution (kinematics) to rate of dominance of
degrees of freedom of a given chirality over their recessive
counterparts. We used the fact that parity is a \ $\mathbb{Z}_2$ \
symmetry to construct the model cube of the argument and used an
isoperimetric inequality to determine that the rate of dominance
has a Gaussian behavior for a large number of degrees of freedom.
We observe that the same kinematic argument can be applied in
determining the phase-space contribution to the rate of
approaching equilibrium for any statistical system possessing a
degree of freedom that can take two possible values. Although, for
concreteness, we used the parity violation as an example in this
paper, the argument can be repeated to any such two-state systems
as well. \\

The spirit of the argument leading to (19) can be traced back to
Boltzmann [27]. It was reached by him in an attempt to derive
statistically the thermodynamic properties of classical ideal
gases from their constituent properties. It was subsequently
conjectured by him to cover all systems, as long as they obey the
ergodic hypothesis [27]. Consequently, it lies at the heart of the
equality of the averages obtained by using either the
micro-canonical or the canonical ensemble in deriving the
thermodynamic properties of equilibrium systems from their
microscopic or mesoscopic constituents [27]. This argument is also
extensively used in derivations involving noise in Langevin-type
equations, especially when reexpressing the stochastic dynamics
through the functional formalism [3],[4]. P. L\'{e}vy derived
these results in a probabilistic setting [22],[23],[26].
Subsequently, they were used extensively in [22],[23] in a
functional analytic and in [26], in a  geometric context. The
present work lies at the confluence of some of these ideas and
also provides a direct application of isoperimetric inequalities
[31], which are used
very extensively in geometry, in a more physically relevant context.\\

Although it may be possible to derive (8),(9) using better-known
analytical techniques, we believe that the method that we follow
here may enhance our intuition about the underlying reasons
determining the rate of convergence of a system toward its
equilibrium state. Moreover, due to the relative insensitivity of
the isoperimetric inequalities [31] and the concentration of
measure phenomenon [22],[23] to the degree of smoothness of the
functions employed, as long as they are Lipschitz, the present
approach may also be applicable in instances where the more
conventional analytic methods may conceivably not be as effective.\\


                                   \vspace{0mm}

\noindent {\small\sc Acknowledgement:} We are grateful to the
referee for his illuminating comments. This research is
supported, in part, by the PSC-CUNY grant 68165-00 37.\\


                               \vspace{3mm}

\centerline{\sc References}

                               \vspace{1mm}

\noindent [1] \ L.D. Landau, E.M. Lifshitz, \ \emph{Statistical Physics, Part 1}, 3rd Ed., \ Pergamon Press,\\
                      \hspace*{5.5mm} Oxford (1980)\\
\noindent [2] \ N.G. van Kampen, \ \emph{Stochastic Processes in Physics and Chemistry}, \ North Holland,\\
                      \hspace*{5.5mm} Amsterdam (1992)\\
\noindent [3] \ J. Zinn-Justin, \ \emph{Quantum Field Theory and Critical Phenomena}, \ 3rd Ed., \ Clarendon\\
                      \hspace*{5.5mm} Press, Oxford (1996)\\
\noindent [4] \ H. Kleinert, \ \emph{Path Integrals in Quantum Mechanics, Statistics, Polymer Physics and\\
                      \hspace*{5.5mm} Financial Markets}, \ 3rd Ed., \ World Scientific, Singapore (2004)\\
\noindent [5] \ \emph{Nonextensive entropy}, \ M. Gell-Mann, C. Tsallis, \ Eds., \ Oxford Univ. Press,\\
                          \hspace*{5.5mm} Oxford (2004)\\
\noindent [6] \ C. Beck, \ {\sf arXiv:cond-mat/0502306}\\
\noindent [7] \ D. Ruelle, \ Proc. Nat. Acad. Sci. USA {\bf 100}, \ 3054, \ (2003)\\
\noindent [8] \ G. Gallavotti, \ {\sf arXiv:cond-mat/0510027}\\
\noindent [9] \ P. Ramond, \ \emph{Journeys Beyond the Standard Model}, \ Perseus Books, Cambridge MA, \\
                      \hspace*{5.5mm} (1999)\\
\noindent [10]  M.J. Ramsey-Musolf, S.A. Page, \ {\sf arXiv:hep-ph/0601127}\\
\noindent [11] H. Takahshi, \ {\sf arXiv:hep-ph/0512321}\\
\noindent [12] D. Kharzeev, \ \emph{Phys. Lett. B} {\bf 633}, \ 260, (2006)\\
\noindent [13] S.A. Voloshin, \ \emph{Phys. Rev. C} {\bf 70}, \ 057901, (2004)\\
\noindent [14]  B. Depslanques, \ \emph{Eur. Phys. J. A} {\bf 24}, \ Suppl. 2, \ 171, (2005)\\
\noindent [15] B. R. Holstein, \ \emph{Fizika B} {\bf 14}, \ 165, (2005)\\
\noindent [16] S.-L. Zhu, C.M. Maekawa, B.R. Holstein, M.J.
                           Ramsey-Musolf, U. van Kolck, \ \emph{Nucl. \\
                     \hspace*{6.5mm} Phys. A} {\bf 748},  \ 435, (2005)\\
\noindent [17] A. Randono, \ {\sf arXiv:hep-th/0510001}\\
\noindent [18] A. Perez, C. Rovelli, \ {\sf arXiv:gr-qc/0505081}\\
\noindent [19] L. Friedel, D. Minic, T. Takeuchi,
                       \ \emph{Phys. Rev. D} {\bf 72}, \  104002, \ (2005)\\
\noindent [20] C.J. Horowitz, \ \emph{Eur. Phys. J. A} {\bf 24S2}, \ 167, (2005)\\
\noindent [21] J. Gu\'{e}na, M. Lintz, M.-A. Bouchiat, \ \emph{Mod. Phys. Lett. A} {\bf 20}, \ 375, (2005)\\
\noindent [22] V.D. Milman, G.Schechtman, \emph{Asymptotic Theory of Finite Dimensional Normed \\
                    \hspace*{6.5mm} Spaces}, Lect. Notes Math. {\bf 1200}, Springer-Verlag, (1986)\\
\noindent [23] V.D. Milman, \ Ast\'{e}risque {\bf 157-158}, \ 273, (1988)\\
\noindent [24]  L.H. Harper, \ \emph{J. Comb. Theor.} {\bf 1}, \ 385,  (1966)\\
\noindent [25] D. Amir, V.D. Milman, \ \emph{Israel J. Math.} {\bf 37}, \ 3,  (1980)\\
\noindent [26] M. Gromov, \ \emph{Metric Structures for Riemannian and Non-Riemannian Spaces},\\
                    \hspace*{6.5mm}  Birkhauser, Boston (1999)\\
\noindent [27] G. Gallavotti, \ \emph{Statistical Mechanics: A Short Treatise}, \ Springer-Verlag, Berlin (1999)\\
\noindent [28] L.J. Boya, E.C.G. Sudarshan, T. Tilma, \ \emph{Rep. Math. Phys.} {\bf 52}, \ 401,  (2003)\\
\noindent [29] E.T. Whittaker, G.N. Watson, \ \emph{A Course of Modern Analysis}, \ Camb. Univ. Press,\\
                    \hspace*{6.5mm} Cambridge (1927)\\
\noindent [30] R. Latala, \ \emph{Conv. Geom. Anal.} {\bf 34}, \ 123,  (1998)\\
\noindent [31] I. Chavel, \ \emph{Isoperimetric Inequalities: Differential Geometric and Analytic \\
                    \hspace*{6.5mm} Perspectives}, \ Camb. Univ. Press, Cambridge (2001)\\

                               \vfill

\end{document}